\documentclass[12pt]{article}
\usepackage[dvips]{graphicx}
\usepackage{pdproc}
\usepackage{amsmath,amssymb,amsfonts,amscd,bbm,graphicx,cite}

  \makeatletter 
  \def\@cite#1{[#1]} 
  \makeatother    
\begin{document}

\renewcommand{\thefootnote}{\alph{footnote}}

\title{
 Higgsless Standard Model in Six Dimensions\footnote{Talk presented at
 {\it SUSY04: The 12th International Conference on Supersymmetry and Unification of Fundamental Interactions}, held at Epochal Tsukuba, Tsukuba, Japan, June 17--23, 2004.}}

\author{ GERHART SEIDL}

\address{ 
Department of Physics, Oklahoma State University, \\
Stillwater, OK  74078, USA
%%%%% You may comment out the e-mail address line below.  
\\ {\rm E-mail: gseidl@susygut.phy.okstate.edu}}

\abstract{
We present a Higgsless Standard Model in six dimensions, based on the
Standard Model gauge group $SU(2)_L\times U(1)_Y$, with two flat extra dimensions compactified on a rectangle. The electroweak symmetry is broken by (mixed)
boundary conditions and realistic gauge boson masses can be accommodated
by proper choice of the compactification scales and brane kinetic terms.
With respect to ``oblique'' corrections, the agreement with
electroweak precision tests is somewhat improved compared to
the simplest five-dimensional Higgsless models.}

\normalsize\baselineskip=15pt

\section{Introduction}

Recently, a new class of Higgsless models has been proposed,
in which electroweak symmetry breaking (EWSB) is accomplished without the Higgs
mechanism by employing mixed boundary conditions (BC's) on a compact space
\cite{Csaki:2003dt,Csaki:2003zu,Barbieri:2003pr}.
These Higgsless models describe a five-dimensional (5D) $SU(2)_L\times SU(2)_R\times U(1)_{B-L}$ gauge theory compactified on an
interval $[0,\pi R]$, where tree-level unitarity of longitudinal gauge boson scattering is ensured through the exchange of the attendant Kaluza-Klein (KK)
tower of massive gauge boson excitations \cite{SekharChivukula:2001hz}.
In a flat extra dimension, four-dimensional (4D)
brane kinetic terms are necessary to decouple at low energies the higher KK
excitations \cite{Barbieri:2003pr}, whereas in Higgsless warped space models they are required \cite{Davoudiasl:2003me} to evade
disagreement with electroweak precision tests (EWPT) \cite{LEP} due to
tree-level ``oblique'' corrections \cite{Peskin:1991sw,Altarelli:1990zd,Holdom:1990tc}.

In this talk, which is based on work done in collaboration with Steven Gabriel and Satya Nandi \cite{Gabriel:2004ua}, we consider a six-dimensional (6D) Higgsless model using only the Standard Model (SM) gauge group $SU(2)_L\times U(1)_Y$.
The model is formulated in flat space with the two extra dimensions compactified on a rectangle and EWSB is achieved by imposing BC's consistent with the variation of the action. The higher KK
resonances of $W^\pm$ and $Z$ decouple below $\sim 1{\rm TeV}$ through the
presence of dominant 4D brane kinetic terms. The
$\rho$ parameter can be set exactly to one by an appropriate choice of the
bulk gauge couplings and compactification scales. Unlike in the 5D theory, the mass scale of the lightest gauge bosons $W$ and $Z$ is set by the dimensionful bulk couplings alone, which are of the order $\sim 10^2\:{\rm GeV}$. Here,
the tree-level oblique corrections to EWPT are somewhat in
better agreement with data than in the simplest 5D warped and flat Higgsless
models.

\section{The Model}

\begin{figure}[htb]
\begin{center}
\includegraphics*[bb = 163 588 449 752]{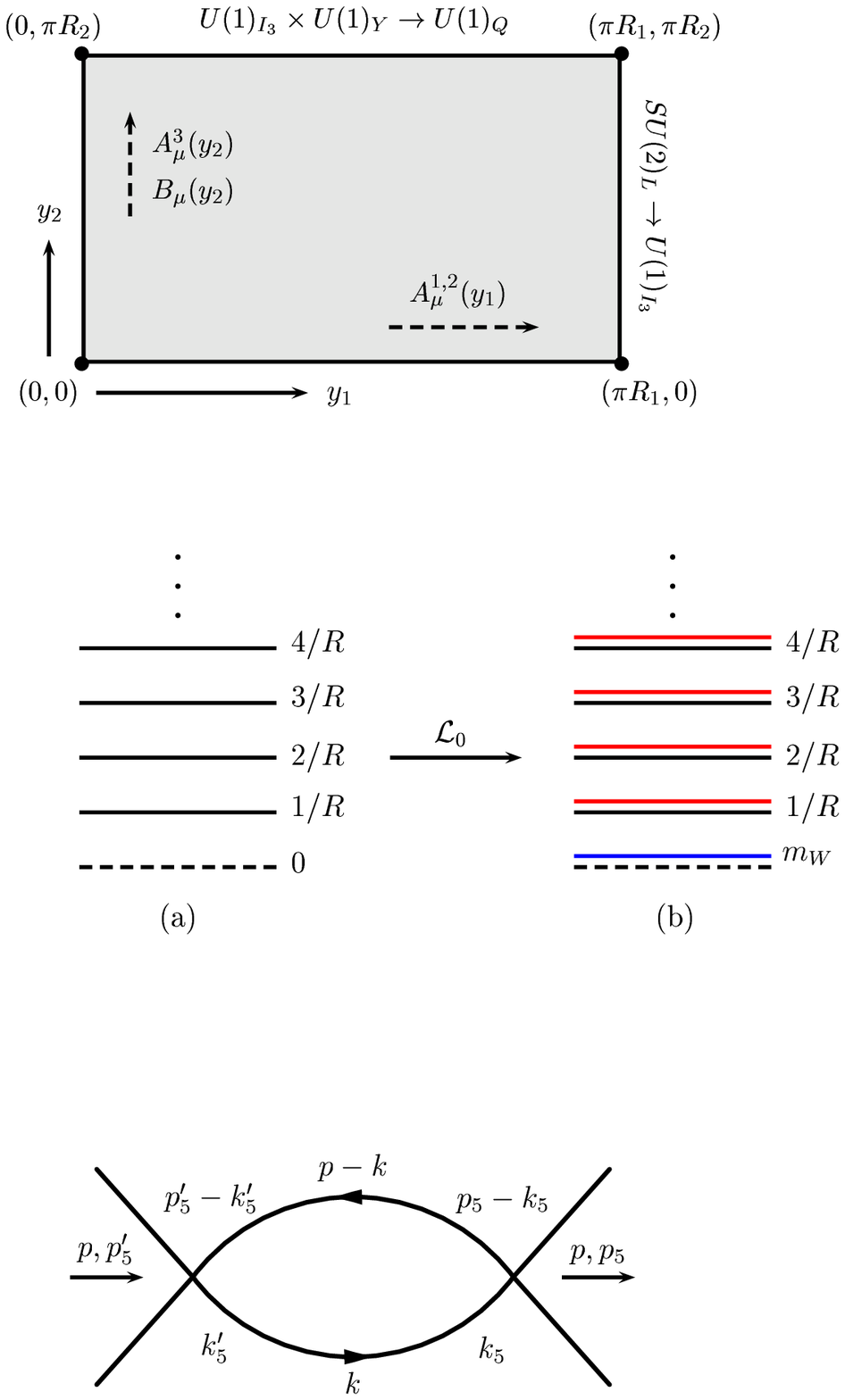}
\caption{%
Symmetry breaking of $SU(2)_L\times U(1)_Y$ on the rectangle.
At one boundary $y_1=\pi R_1$, $SU(2)_L$ is broken to $U(1)_{I_3}$ while
on the boundary $y_2=\pi R_2$ the subgroup
$U(1)_{I_3}\times U(1)_Y$ is broken to $U(1)_Q$ of electromagnetism, which leaves
only $U(1)_Q$ unbroken on the entire rectangle.
}
\label{fig:rectangle}
\end{center}
\end{figure}

Consider a 6D $SU(2)_L\times U(1)_Y$ gauge theory in a flat space-time background, where the two extra spatial dimensions are compactified on a rectangle \cite{Gabriel:2004ua}. If we denote by $y_1$ and $y_2$ the coordinates of the 5th and 6th dimension, the physical space is defined by
$0\leq y_1\leq \pi R_1$ and $0\leq y_2\leq\pi R_2$. The $SU(2)_L$ and $U(1)_Y$ gauge bosons in the bulk are respectively written as
$A_M^a$ ($a=1,2,3$ is the gauge index) and $B_M$, where capital Roman letters
$M=0,1,2,3,4,5,6$ denote the 6D Lorentz indices, while Greek letters $\mu=0,1,2,3$ symbolize the usual 4D Lorentz indices. The action of the gauge fields in our model is given by
\begin{equation}\label{eq:S}
 \mathcal{S}=\int d^4x\int_0^{\pi R_1}dy_1\int_0^{\pi R_2}
dy_2\left(\mathcal{L}_6+\delta(y_1)\delta(y_2)\mathcal{L}_{0}\right),
\end{equation}
where $\mathcal{L}_6$ is a 6D bulk gauge kinetic term and
$\mathcal{L}_0$ is a 4D brane gauge kinetic term localized at
$(y_1,y_2)=(0,0)$, which read respectively
\begin{equation}\label{eq:gaugekinetic}
 \mathcal{L}_6  = 
 -\frac{M_L^2}{4}F^a_{MN}F^{MNa}-\frac{M_Y^2}{4}B_{MN}B^{MN},\quad
 \mathcal{L}_0  =  
 -\frac{1}{4g^2}F_{\mu\nu}^a F^{\mu\nu a}-\frac{1}{4{g'}^2}B_{\mu\nu}
B^{\mu\nu},
\end{equation}
with field strengths
$F_{MN}^a=\partial_M A^a_N-\partial_N A^a_M+f^{abc}A^b_MA^c_N$
($f^{abc}$ is the structure
constant) and $B_{MN}=\partial_MB_N-\partial_N B_M$. In Eqs.~(\ref{eq:gaugekinetic}), the quantities $M_L$ and $M_Y$ have mass dimension $+1$, while
$g$ and $g'$ are dimensionless. Now, EWSB $SU(2)_L\times U(1)_Y\rightarrow U(1)_Q$ is achieved by imposing suitable Dirichlet, Neumann, and mixed BC's \cite{Gabriel:2004ua}, which are consistent with the variation of the action and correspond therefore to a soft
gauge symmetry breaking. Schematically,
the symmetry breaking is sketched in Fig.~\ref{fig:rectangle}.
The fermions are, like the gauge bosons, approximately localized by dominant brane kinetic terms at $(y_1,y_2)=(0,0)$, thus suppressing for the light generations unwanted non-oblique corrections to the electroweak precision parameters.

The total effective 4D Lagrangian in the compactified theory $\mathcal{L}_{\rm total}$ can be
written as $\mathcal{L}_{\rm total}=\mathcal{L}_0+\mathcal{L}_{\rm eff}$, where
$\mathcal{L}_{\rm eff}=\int_{0}^{\pi R_1}
dy_1\int_0^{\pi R_2}dy_2\:\mathcal{L}_6$ denotes the contribution from the
bulk, which follows from integrating out the extra dimensions. Here,
$\mathcal{L}_{\rm eff}$ generates electroweak vacuum polarization amplitudes summarizing in the 4D theory the effect of the symmetry breaking sector. These
vacuum polarizations lead at tree-level to oblique corrections (as opposed to vertex corrections and box diagrams) of the gauge boson propagators and thus
affect electroweak precision measurements \cite{Peskin:1991sw,Altarelli:1990zd}. To determine the KK masses of the gauge bosons, we will from now on assume
that the brane terms $\mathcal{L}_0$ dominate the bulk kinetic terms,
{\it i.e.}, we take $1/g^2,1/{g'}^2\gg (M_{L,Y}\pi)^2R_1R_2$. As a result, we
find from $\mathcal{L}_{\rm eff}$ for the
$W^{\pm}$'s the mass spectrum
\begin{equation}\label{eq:mW}
 m_0^2\approx 2g^2M_L^2R_2/R_1\:\:=\:\:m_W^2,\quad
 m_n\approx n/R_1,\quad n=1,2,\ldots,
\end{equation}
where we identify the lightest state with mass $m_0$ with the $W^\pm$. Observe
in Eq.~(\ref{eq:mW}), that the inclusion of the brane kinetic terms
$\mathcal{L}_0$ for $1/R_1,1/R_2 \gtrsim
\mathcal{O}({\rm TeV})$ leads to a decoupling of the higher KK-modes with masses
$m_n$ $(n>0)$ from the electroweak scale, leaving only the $W^\pm$ states
with a small mass $m_0$ in the low-energy theory. The lowest massive state in the tower of the neutral gauge bosons has a mass-squared
\begin{equation}\label{eq:mZ}
 m_Z^2\approx2(g^2+{g'}^2)M_L^2M_Y^2 R_1/[(M_L^2+M_Y^2)R_2]
\end{equation}
which we identify with the $Z$ of the SM. All other KK modes of the
$\gamma$ and $Z$ have masses of order $\gtrsim 1/R_2$ and thus decouple for
$1/R_1,1/R_2\gtrsim\mathcal{O}({\rm TeV})$, leaving only
a massless $\gamma$ and a $Z$ with mass $m_Z$ in the low-energy
theory.

\section{Relation to EWPT}\label{sec:EWPT}
One crucial test for any model of EWSB is the value of the $\rho$ parameter,
which is experimentally known to satisfy
the relation $\rho=1$ to better than 1\%. In our model, we choose
the 4D brane couplings $g$ and $g'$ to follow the usual SM relation
$g^2/(g^2+{g'}^2)={\rm cos}^2\theta_W\approx 0.77$. Defining
$\rho=1+\Delta\rho$, we then obtain that
$\Delta\rho =0$ if the bulk kinetic couplings and compactification radii
satisfy the relation $(M_L^2+M_Y^2)/M_Y^2=R_1^2/R_2^2$.
Although we can thus fit $\Delta\rho=0$ by appropriately dialing the model
parameters, $\mathcal{L}_{\rm eff}$ 
introduces a manifest breaking of custodial symmetry
and will thus contribute to EWPT via oblique corrections to the SM
parameters. The effects of oblique corrections on EWPT can be parameterized
in the $\epsilon_1$, $\epsilon_2$, and $\epsilon_3$ framework
\cite{Altarelli:1990zd}, where the current experimental bounds on the
relative shifts with respect to the SM expectations are roughly of the order
$\epsilon_1,\epsilon_2,\epsilon_3\lesssim 3\cdot 10^{-3}$ \cite{Barbieri:2004qk}. For our choice of parameters, we consistently find
$\epsilon_1=\Delta\rho=0$. The quantities $|\epsilon_2|$ and
$|\epsilon_3|$, on the other hand, are bounded from below by the requirement of
having sufficiently many KK modes below the strong coupling (or cutoff)
scale $\Lambda$ of the theory. In the 6D model, we would naively estimate
$\Lambda\simeq\sqrt{2}(4\pi)^{3/2}M_{L,Y}$ \cite{Chacko:1999hg} which leads for
$M_{L,Y}\simeq 10^{2}{\rm GeV}$ to $\Lambda\simeq 6\:{\rm TeV}$. Assuming $M_L=M_Y$, we have $R_2=R_1/\sqrt{2}$ and
\begin{eqnarray}\label{eq:epsilon3value}
 \epsilon_3\simeq\frac{g^2}{96\sqrt{2}\pi}(\Lambda R_2)^2\simeq
  2.3 \times10^{-3}
\times (g\Lambda R_2)^2,
\end{eqnarray}
while $\epsilon_2\simeq\epsilon_3$. It is interesting to compare Eq.~(\ref{eq:epsilon3value}) with the corresponding result of the 5D model in Ref.~\cite{Barbieri:2003pr}. We find that the parameter $\epsilon_3$ is in the 6D model by $\sim 15\%$ smaller than the
corresponding 5D value.
This is due to the fact that in the 6D model the bulk gauge kinetic couplings satisfy
$M_L=M_Y\simeq 100\:{\rm GeV}$, while they take in 5D only the values
$M_L\simeq M_Y\simeq 10\:{\rm GeV}$. From
Eq.~(\ref{eq:epsilon3value}) we then conclude that the inverse loop
expansion parameter can be $\Lambda R_2\simeq1/g\approx 1.6$ in agreement
with EWPT. Like in the 5D case, however, the 6D model seems not to
admit a loop expansion parameter in the regime $\Lambda R_2\gg 1$ as required
for the model to be calculable.

To summarize, we have considered a 6D Higgsless Standard Model in compactified flat space, which is based on the gauge group $SU(2)_L\times U(1)_Y$. Dominant brane interactions lead to a realistic gauge sector and the model parameters
allow to improve, with respect to oblique corrections, the fit of 
EWPT as compared to the simplest 5D Higgsless models.

\section{Acknowledgements}

I would like to thank my collaborators Steven Gabriel and Satya Nandi.
This work is supported by the U.S. Department of Energy under
Grant Numbers DE-FG02-04ER46140 and DE-FG02-04ER41306.

\bibliographystyle{plain}

\end{document}